\documentclass[preprint]{revtex4}
\usepackage{graphics,graphicx,epsfig}

\begin{document}

\title{Thin-film growth by random deposition of rod-like particles on a square lattice}

\author{F. L. Forgerini}
\email{fabricio_forgerini@ufam.edu.br}
\affiliation{ISB, Universidade Federal do Amazonas, 69460-000 Coari-AM, Brazil and \\
Departamento de F\'{\i}sica, I3N - Universidade de Aveiro, 3810-193 Aveiro, Portugal}

\author{W. Figueiredo}
\email{wagner@fisica.ufsc.br}
\affiliation{Departamento de F\'{i}sica, Universidade Federal de Santa Catarina,
88040-900 Florian\'opolis-SC, Brazil }

\date{\today}
\begin{abstract}
Monte Carlo simulations are employed to investigate the surface growth generated by deposition of particles of different sizes on a substrate, in one and two dimensions. The particles have a linear form, and occupy an integer number of cells of the lattice. The results of our simulations have shown that the roughness evolves in time following three different behaviors. The roughness in the initial times behaves as in the random deposition model, with an exponent $\beta_{1} \approx 1/2$. At intermediate times, the surface roughness depends on the system dimensionality and, finally, at long times, it enters into the saturation regime, which is described by the roughness exponent $\alpha$.
The scaling exponents of the model are the same as those predicted by the Villain-Lai-Das Sarma equation for deposition in one dimension. For the deposition in two dimensions, we show that the interface width in the second regime presents an unusual behavior, described by a growing exponent $\beta_{2}$, which depends on the size of the particles added to the substrate. If the linear size of the particle is two, we found that $\beta_{2}<\beta_{1}$, otherwise it is $\beta_{2}>\beta_{1}$, for all particles sizes larger than three. While in one dimension the scaling exponents are the same as those predicted by the Villain-Lai-Das Sarma equation, in two dimensions, the growth exponents are nonuniversal.


\end{abstract}

\maketitle

\section{Introduction}

The field of non-equilibrium statistical physics is related to those problems for which a hamiltonian formulation is not possible or the detailed-balance condition is not satisfied. One of these problems is the study of surface growth and interfaces, where the understanding of the basic mechanisms is the key to developing thin-film devices with important technological applications \cite{barabasi,meakin}.

In this work we are interested in the description of the morphology of a surface formed by adding particles to an initially flat substrate. Particles, having one unit height and linear size $N$, land horizontally onto the surface, or over a straigth line in the case of a one-dimensional substrate, and are not allowed to diffuse. In this way, we calculate the interface width, $w(L,t)$, a function that determines the roughness of the interface, where $L$ is the linear size of the substrate and $t$ is the time variable. The simplest known model for the deposition of particles is the random deposition, where particles are aggregated onto an initially flat substrate, and lateral correlations among the deposited particles are completely neglected. For this trivial model, its continuous and discrete atomistic versions have exact solutions.

The simplest non-trivial growth models, which incorporate spatial and temporal correlations, can be described by means of stochastic differential equations. The Edwards-Wilkinson (EW) equation \cite{EW82} is designed to take into account the surface relaxation and height-height correlations in some deposition models. On the other hand, when we need to include some lateral growth of the interface, as in the ballistic deposition of particles \cite{Vold}, the appropriate stochastic equation is the non-linear Kardar-Parisi-Zhang (KPZ) equation \cite{KPZ86}. The (EW) and (KPZ) equations can be written as

\begin{equation}
\frac{\partial h({\textbf{r}},t)}{\partial t}=\nu\nabla^{2}h+\eta({\textbf{r}},t)\hspace{1cm}[EW]
\label{EW}
\end{equation}
and
\begin{equation}
\frac{\partial h({\textbf{r}},t)}{\partial t}=\nu{\nabla}^{2}h+\frac{\lambda}{2}(\nabla h)^{2}+\eta({\textbf{r}},t) \hspace{1cm} [KPZ],
\label{KPZ}
\end{equation}
where $\eta({\textbf{r}},t)$ represents the random fluctuations in the deposition processes.

In order to calculate the interface width, we determine the vertical height of a given point of the surface relative to the substrate, $h(\textbf{r},t)$, where \textbf{r} denotes the position of a cell over the substrate. The roughness $w(L,t)$ is defined as the mean square fluctuation of the height, $w(L,t)=\langle[h(\textbf{r},t)-\overline{h}(t)]^{2}\rangle^{1/2}$, where $\overline{h}(t)$ is the average value of the surface height at a given instant of time $t$.

It is well established that for a large class of growth models the Family-Vicsek scaling relation applies \cite{FV}

\begin{equation}
w(L,t) \sim L^{\alpha}f(\frac{t}{L^{z}}),
\label{family_vicsek}
\end{equation}
where the scaling function $f(x)$ is a constant for very large values of $x$, and $f(x) \sim x^{\beta}$ when $x << 1$. The exponent $\alpha$ characterizes the interface width at long times, $z$ is the dynamic exponent, while $\beta$ is the exponent associated to the initial stages of the growth. These exponents are not at all independent, and are related by $\alpha = \beta z$.

In the theoretical studies performed in the area of surface growth, one is interested in the calculation of these exponents. There are many studies in the literature concerning the properties of the growth models where calculations have been done analytically, or through the extensive use of Monte Carlo simulations
\cite{das_sarma,albano,drossel,cavalcanti,Reis,barato_oliveira}.

In this work we use Monte Carlo simulations to study the surface growth in $(1+1)$ and $(2+1)$ dimensions due to the deposition of particles of different sizes. There are in literature some studies where two or more different deposition models are combined \cite{cerdeira_95,elnassar_cerdeira}, or two different species of particles are deposited \cite{caglioti,trojan,karmakar} in order to describe real systems. Here, we choose our particles to be deposited from a modified Poisson distribution. This type of distribution appears to be relevant in some ash particles deposition on the heat exchange surfaces \cite{Baxter,forgerini_figueiredo}.

Recently, we have investigated the surface growth generated by the random deposition of particles of different sizes in one and two dimensions \cite{forgerini_figueiredo,fw2}. The results of our simulations have shown that in both dimensions, the roughness evolves in time following three different behaviors. The roughness at the initial times behaves as in the random deposition model, where correlations are absent. At intermediate times, the surface roughness depends on the dimensionality of the system, and finally, at long times, it enters into the saturation regime. However, there is a remarkable difference in the results we find in one and two dimensions, regarding the behavior of the interface width at the intermediate times. While in one dimension, the growth exponent is independent of the particle size, in two dimensions it increases with the size of the particle. This happens because when we try to deposit a new particle in one dimension, it can be accomodated in one of the two ends of an already deposited particle, independently of its size. However, in two dimensions, the new deposition can be done in the $2(1+N)$ positions around a deposited particle, where $N$ is the linear size of the particle. This dependence on $N$ makes the growth in two dimensions to be non-universal.

\section{Model description}

\begin{figure}[htb]%
\includegraphics*[scale=0.6]{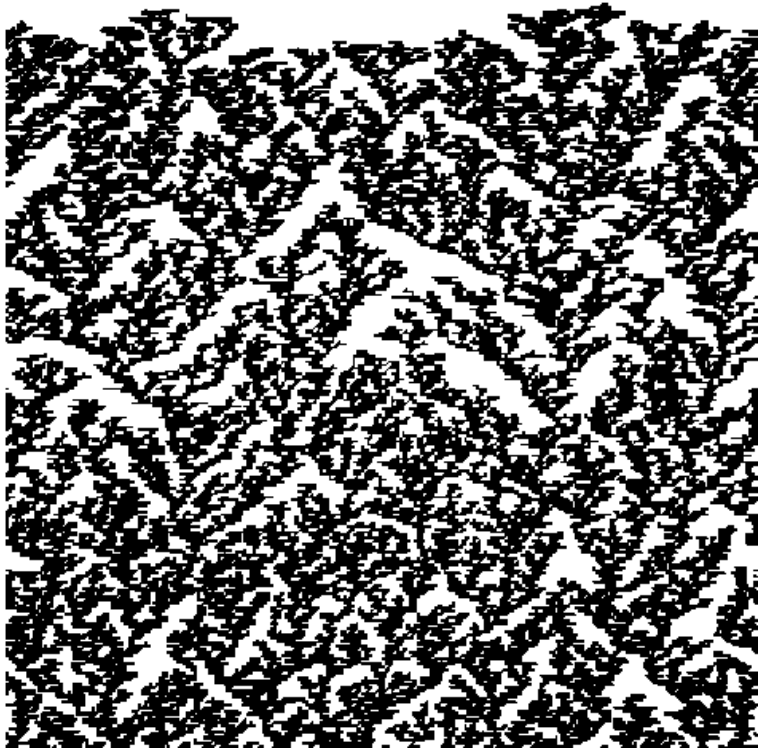}
\caption{Graphic representation of the deposition in a one-dimensional lattice of linear size $L=256$. In this figure particles are selected from a modified Poisson distribution of mean value $N=5$. Time grows from bottom to top, where the interface width reached the stationary state.}
\label{deposition}
\end{figure}

In the present study particles of linear size $N$ are dropped randomly over a finite one- or two-dimensional substrate. All the particles considered in the deposition process are one unit height and length sizes in the range $1\leq N\leq15$. The most important difference between this model and the simple random deposition model, is that our model naturally allows for correlations among the columns, that is, it leads to a lateral growth of the interface. Figure~\ref{deposition} is a graphic representation of the one-dimensional deposition of a mixture of particles, which are selected from a modified Poisson distribution, where the maximum particle size is $N=15$.

In two dimensions, a particle is aggregated only if the chosen site for deposition coincides with the midpoint of the particle, and there is enough space to accomodate it in the $x$ or $y$ directions. Otherwise, if the cell is already occupied or there is no enough space for the deposition, the particle is reflected away due to these geometric constraints. Particles are not allowed to diffuse or share the same cell in any deposition plane. When we are depositing identical particles, during a unit of time, we try to deposit $L^{2}/N$ particles, which roughly means the deposition of one layer. For the deposition of particles of same size $N$ in a one-dimensional substrate, during a unit of time we try to deposit $L/N$ particles. Therefore, depending on the size of the particle, many trials of deposition are not successful in the unit of time.

\section{Results}
\label{results}

\begin{figure}[ht]
  \centering
  \vspace{0.4cm}
  \includegraphics[scale=0.4]{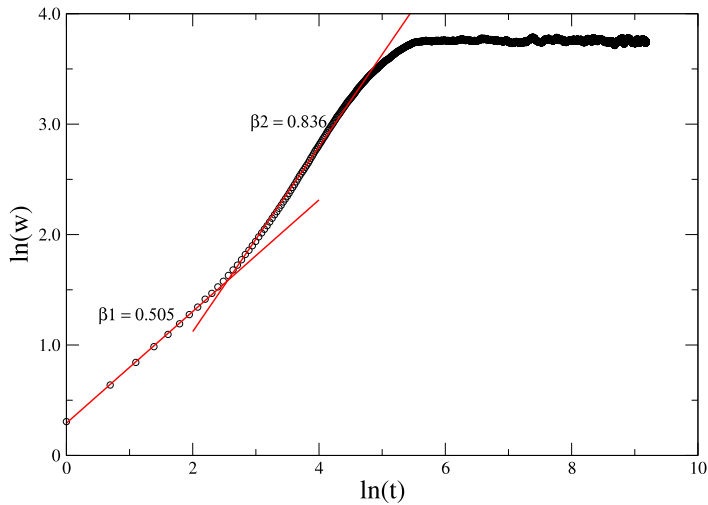}
  \caption{(Color online) Log-log plot of the interface width versus time for a two-dimensional substrate. Deposition of particles of size $N=8$, and the linear size of the square lattice is $L=128$.}
  \label{roughnessXt}
\end{figure}

In this section we only show some plots for the deposition in two dimensions. Our Monte Carlo simulations were performed on square lattices with linear size ranging from $L=32$ to $512$. We exhibit in Figure~\ref{roughnessXt} the log-log plot of the interface width as a function of time for particles of size $N=8$. At the initial times of deposition, the growth exponent $\beta_{1}$ is close to that of the random deposition. At intermediate times, the growth exponent exhibits a completely different behavior from the one observed in one dimension \cite{forgerini_figueiredo,fabio_2,fabio_silveira}. In one dimension, the growth exponent $\beta_{2}$ at intermediate times, does not depend on the size of particle and its value is $\beta_{2}=0.31$. In two dimensions, this exponent exhibits a very peculiar behavior. While for $N=2$ its value is around $0.20$, which is smaller than $\beta_{1}$, it jumps, for $N=3$, to the value $0.63$. For $N > 3$ it increases, but the rate of increasing is lower for higher values of $N$.

\begin{figure}[htb]
\vspace{0.4cm}
\includegraphics[scale=0.4]{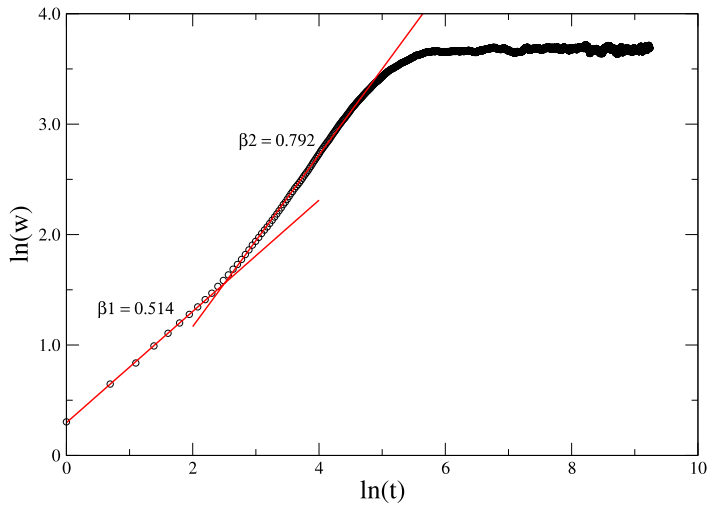}
\caption{(Color online) Log-log plot of the interface width for the deposition of particles with sizes $1\leq N\leq15$ on a square lattice of side $L=128$. The values of the growth exponents $\beta_{1}$ and $\beta_{2}$ are indicated in the plots.}
\label{roughness}
\end{figure}

We show in Figure~\ref{roughness} the behavior of the interface width for the deposition on a square substrate of linear size $L=128$ for a mixture of particles with sizes in the range $1\leq N\leq15$. Every time we try to deposit a new particle, it is selected from a modified discrete Poisson distribution whose maximum value is $N=15$. As to be expected the figures we find for the scaling exponents are intermediary between those calculated for the smallest and largest particles.

In Figure 4 we display the values of the growth exponent $\beta_{2}$ as a function of $N$ for three different lattice sizes. We see that it is almost independent of $L$, but the rate of increasing is very small for higher values of $N$. Although not shown, we have also calculated the roughness exponent $\alpha$ by extrapolating the saturation value of the interface width observed at long times for large values of $L$ and different particles sizes. We have found that $\alpha$ is very close to $1/3$. For the deposition on a linear substrate we have determined the value $\alpha = 0.94$, which is also independent of $N$. Therefore, in one dimension, using this value of $\alpha$, along with $\beta_{2}=0.31$, the Family-Vicsek relation gives directly the value $z=3.0$, for any value of particle length $N$. These figures put the one-dimensional version of this model in the same universality class of the Villain-Lai-Das Sarma equation \cite{LaiSarma}. Despite $\alpha$ is close to $1/3$ in two dimensions, the Family-Vicsek relation predicts values for the dynamic exponent $z$ decreasing with the size $N$ of the particles. Then, the behavior of the model in two dimensions is nonuniversal.

\begin{figure}[htb]
\vspace{0.4cm}
 \includegraphics[scale=0.4]{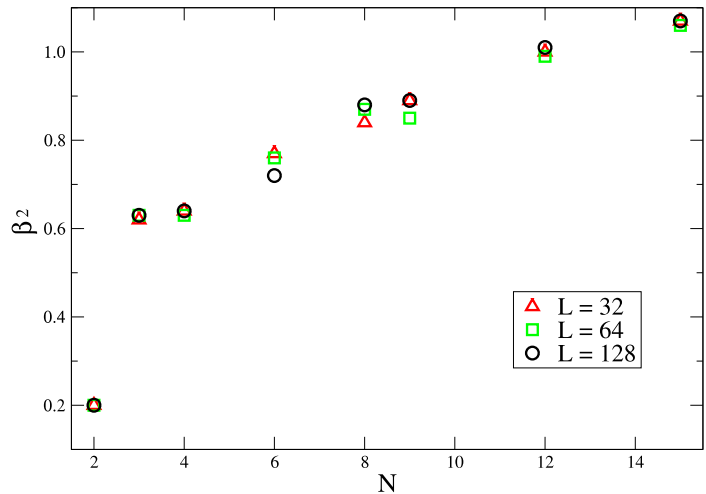}\\
\caption{(Color online) Plot of the growth exponent $\beta_{2}$ as a function of the particle length $N$ for three different linear sizes $L$ of the square lattice as indicated in the figure.}
 \label{fig:roughness}
\end{figure}

\section{Conclusions}
\label{conclusions}

We employed Monte Carlo simulations to study the surface growth due to the deposition of rod-like particles of different sizes over linear and square lattices.We have shown that the interface width as a function of time exhibits three different behaviors. At the initial times its behavior is similar to the usual random deposition model. At intermediate times, the surface roughness depends on the system dimensionality and finally, at long times, it saturates. We have seen that in one dimension the dynamics of the model can be mimicked by the Villain-Lai-Das Sarma equation for whatever particle size. On the other hand, for the deposition in two dimensions, the growth exponent at intermediary times changes with the particle size and the deposition model behaves differently from one dimension. For the deposition in two dimensions, we show that the interface width in the second regime presents an unusual behavior, described by a growth exponent $\beta_{2}$, which depends on the size of the particles added to the substrate. If the linear size of the particle is two, we found that $\beta_{2}<\beta_{1}$, otherwise it is $\beta_{2}>\beta_{1}$, for all particles sizes larger than three. While in one dimension the scaling exponents are the same as those predicted by the Villain-Lai-Das Sarma equation, in two dimensions, the growth exponents are nonuniversal.

\begin{acknowledgements}
The authors would like to thank Conselho Nacional de Desenvolvimento Cient\'{i}fico e Tecnol\'ogico (CNPq) for the financial support.
\end{acknowledgements}

\end{document}